# Can Astronomy Manage Its Data?

Ray Norris

*Vice-President, IAU Commission 5 and IAU delegate to CODATA*

## Abstract

Astronomy has a distinguished tradition of using technology to accelerate the quality and effectiveness of science, and data-intensive initiatives such as the Virtual Observatory lead the way amongst other fields of science. However, astronomical data are not uniformly well-managed, and our current freedom to create open-access databases is threatened by those who would like all data to be subject to strict Intellectual Property controls. We, like other fields of science, need to establish and agree on a set of guiding principles for the management of astronomical data.

## 1. Introduction

Astronomy, like many other sciences, is in the midst of a data revolution, which offers both opportunities and threats. On one hand, we see the achievements of data centres like CDS and NED, the vigorous international development of the Virtual Observatory (VO), the revolutionary public data releases from individual astronomical projects, and the rapid dissemination of results made possible by forward-thinking journals, ADS, and astro-ph. All of these position astronomy as a role model to other sciences for how technology can be used to accelerate the quality and effectiveness of our science.

On the other hand, there are international pressures to surround our open-access databases in a morass of legal red tape, and we are poorly prepared to resist these. In some cases, the management and archiving of our valuable terabyte databases are haphazard and left to individual scientists to manage as best as they can. Vital data are published in journals but never make it into the data centres, while other valuable data gather dust in observatories, inaccessible, endangered, and in some cases undigitised. While many of us revel in our rapid electronic access to astronomical databases and journals, our colleagues in developing countries are unable to participate in this revolution, and are in danger of being left behind.

Astronomy needs to address these challenges: first, by establishing a consensus as to what represents good data management practice in astronomy, and, second, by building this consensus into a framework of guiding principles. We can maximise our effectiveness in this task if we work with colleagues across the spectrum of international science, who are facing very similar challenges.

## 2. Threats to Scientific Data

In 2000-2, our freedom to create and use open-access astronomical databases was threatened by legislation proposed by WIPO (World Intellectual Property Organisation), the European Union, and other bodies. Under that legislation, a system of licensing would be established under which any use of data would need to be accompanied by a paper trail to prove that use of the data was legitimate. Unlike the current copyright laws, which generally work well, there would be no provision for "fair use" for education or research. This would effectively make unworkable our current practice of freely distributing data and information.

While most practising scientists were comfortably unaware that a war was being fought on their behalf, this threat was beaten off in a series of hard-won legal and political battles by members of ICSU[1] and CODATA[2]. With hindsight, the problem arose because science had never articulated any broad data management principles or policies. As a result, organisations concerned with Intellectual Property Rights had erroneously concluded that science had no opinion on how data should be managed, leaving them free to impose their views and legislation on scientific data.

A lesson from this is that the science community must ensure that its data needs are better articulated and understood, emphasising, for example, the legitimacy of open-access databases. However, some of the criticisms levelled at science were, sadly, justified. For example, it is sometimes arbitrary and haphazard as to whether valuable data obtained at taxpayers' expense is ever made available to the rest of the scientific community, or is adequately protected and archived.

In the face of increasing pressures to place legal and commercial restrictions on access to data and information, the scientific community has recognised the need to establish sound data management principles. In early 2004, the ICSU set up a panel of independent experts to perform a "Priority Area Assessment on Scientific Data and Information". The resulting report is comprehensive and visionary, and proposes a way forward for the scientific community to create a new global framework for data and information policy and management. The IAU and other scientific Unions are encouraged to participate in this. The IAU is well-represented in this by its active participation in CODATA, and this paper is part of that process.

Other important initiatives include the World Summit on the Information Society (WSIS), which has been endorsed by the United Nations General Assembly, and which has declared, amongst other things, that "We strive to promote universal access with equal opportunities for all to scientific knowledge and the creation and dissemination of scientific and technical information, including open access initiatives for scientific publishing."

Furthermore, in January 2004 the OECD[3] made a "Declaration on Access to Research Data from Public Funding" which essentially states that the governments concerned (representing the bulk of IAU membership) will work towards making publicly-funded data openly accessible.

## 3 Other Issues in Astronomical Data Management

### *3.1 Open Access*
Because the advance of astronomy frequently depends on the comparison and merging of disparate data, it is important that astronomers have access to all available data on the objects or phenomena that they are studying. Astronomical data have therefore always enjoyed a tradition of open access, best exemplified by the astronomical data centres, which provide access to data for all astronomers at no charge.

---

[1] ICSU is the International Council of Science, the peak organisation to which the IAU and other Scientific Unions subscribe.
[2] CODATA is the Committee for Scientific Data of the ICSU
[3] OECD is the Organisation for Economic Cooperation and Development, and the full statement may be found on http://www.oecd.org/document/0,2340,en_2649_34487_25998799_1_1_1_1,00.html

There exist a number of exceptions to this open access tradition, some of which are widely-supported, such as the initial protection of observers' data by national facilities. However, in a few cases, observatories allow data archive access only to affiliated scientists, while still benefiting from the open access policies of other institutions. This asymmetry continues to be a cause for concern in astronomy.

At the 2003 IAU General Assembly a resolution was adopted that, broadly, says that publicly-funded archive data should be made available to all astronomers. This is aligned with ICSU and OECD recommendations, and may be regarded as a first step towards a articulating the principles by which the astronomical community would like to see its data managed.

### *3.2 Crossing the Digital Divide*

The "Digital Divide" refers to the widening gulf between those with high-bandwidth access to information, data, and web services, and those who do not. Those who do not are further disadvantaged by this lack of access, making it even less likely that they will gain access in the future. Often the term is used to refer to the gulf between developing and developed nations, but it can also refer to the poor information services available to indigenous (and often geographically remote) inhabitants of an otherwise affluent developed country.

Astronomers in developing countries are better positioned than their colleagues in some other disciplines, because most astronomical information is already accessible to them, provided they have adequate internet bandwidth. However, there remain challenges, such as electronic access to journals that are funded by subscriptions from their users, and access by those with poor connectivity, which have yet to be addressed satisfactorily.

### *3.3 Formats for data in Journals*

Since most astronomical journals now publish in electronic form, it might be expected that all data-rich articles would automatically enter the archives of the major astronomical data centres. However, H. Andernach has shown (private communication) that in a representative sample of the electronic tables of 1500 such articles collated from the literature, less than 50% appear in the archive of a data centre. The problems of making accessible this collection, and eventually the data content of the entire astronomical literature, are:
- lack of manpower for writing metadata,
- non-standard data formats in the electronic publication,
- inadequate nomenclature for astronomical objects used by authors,
- the need to scan articles from older (non-electronic) journals.

To make the data in journals available to the data centres and the virtual observatory, standards for presentation of tables in journals need to be established and adhered to. While some journals are already making excellent progress in this direction, others are not. This task will be amongst those addressed by the data framework discussed below.

### *3.4 Preservation and Digitisation of Photographic Plates*

Astronomy possesses a large and potentially valuable reserve of heritage data - about 3 million photographic observations - that have accumulated in plate archives since the late 19th century. However, only a few of those data are accessible in digital form, and most are therefore inaccessible to most potential users. Little attention has been paid to the salvage of historic material, and expertise and equipment have become lost. Plate archives face loss and deterioration through ageing, disasters, and ignorant destruction. Resources for rescuing these data are necessarily in competition with those required to generate new data, and so it is important to determine what value to place on the historical archives, before we lose the opportunity to make that decision.

## 4. A Strategic Framework for Managing Astronomical Data

As the size and complexity of our databases increase, the management of these data must be taken increasingly seriously. However, some major projects are still being conceived and funded without any serious thought being given as to how the data will be managed.

Several very vigorous and effective groups in astronomy (e.g. the data centres, and some major observatories) are individually achieving ambitious goals in the area of data management and handling. However, between and outside these active groups are gaps in which data management is neglected or dealt with in an *ad hoc* way. Whilst the VO is attempting to make major databases accessible to all astronomers, it cannot do so unless those databases are properly constructed and managed. Astronomy does not have any strategic data framework that links these activities together, provides policies or guidelines for astronomical data management, or is able to represent the interests of astronomical data management to external parties. As a result
- We are vulnerable to external threats such as the WIPO legislation described above,
- We are unable to represent astronomical data requirements in a coordinated way to external groups, such as ICSU, funding agencies, or journal publishers,
- There is no uniform approach across astronomy to preservation and dissemination of data,
- While some groups in astronomy adopt a professional approach to data management, others treat it as an afterthought, or neglect it completely, so that astronomy as a whole loses value,
- There is poor coordination between astronomy and other disciplines, and poor recognition in other disciplines of the data needs and strengths of astronomy.

It is important that the astronomy community is able to agree on a set of principles for managing data, so that we can:
- Help funding agencies and institutions understand the issues and principles involved in astronomical data management,
- Combat threats such as the WIPO legislation,
- Train the next generation of scientists in professional data management techniques,
- Address the issues described in Section 3 above,
- Maximise the science that we generate from our data.

Similar issues exist throughout all of science, and the ICSU's "Priority Area Assessment" described above recommends that a "Framework" of such principles needs to be established across all sciences. By setting up a framework focussed on astronomy, the IAU can ensure that its needs are addressed by the ICSU document.

IAU Commission 5 therefore proposes to develop a strategic framework for data management in astronomy, with recommendations to guide and assist individual observatories and organisations, and encouraging principles of open access as far as possible. It will do so in close liaison with the IVOA[4], which can provide the tools and infrastructure for facilitating this process. Recognising that the ICSU and CODATA are also engaging in a similar activity across all sciences, the IAU will actively work with ICSU and CODATA, both to participate in the ICSU framework, and to bring that experience to the development of an astronomical framework.

## 5. Conclusion

To safeguard the future of astronomical data, and to ensure that astronomers are able to reap the maximum scientific benefit from their data, it is essential that the astronomy community agree on a set of principles by which astronomical data should be managed. To achieve this goal, IAU Commission 5 representatives will:

- Actively participate in ICSU and CODATA discussions
- Conduct electronic discussions within IAU to reach broad agreement on the way forward
- Collaborate closely with the IVOA on developing requirements for implementing these strategies within the VO
- Hold an open meeting at the Prague IAU GA in 2006, at which a draft framework will be debated
- Propose a resolution at the IAU GA in 2006 for the IAU to adopt and develop the data management framework.

The ICSU report presents a vision of scientific data management that is well-aligned with the strategic goals of astronomy. The astronomical community should embrace the opportunity to make its data management strategy explicit, and work with colleagues in other scientific disciplines to bring sound data management principles to astronomy. To do so will not only further astronomy, but could generate a vigorous growth in the availability and inter-operability of astronomical data, resulting in even more cross-fertilisation and delivery of cutting-edge science.

## Acknowledgements

I thank my colleagues in the IAU Working Group for Astronomical Data, in IAU Commission 5, and in CODATA, for their participation in discussions that led to this paper.

---

[4] IVOA is the International Virtual Observatory Alliance